\documentclass[twocolumn,showpacs,preprintnumbers,amsmath,amssymb]{revtex4}

\usepackage{natbib}
\usepackage{lgrind}
\usepackage{latexcad}
\usepackage{float}
\usepackage{amssymb}
\usepackage{amsmath,amsfonts}
\usepackage{graphicx}
\usepackage{rotating}
\usepackage{dcolumn}
\usepackage{mathrsfs}
\usepackage{bm}

\begin{document}

\baselineskip=5mm


\title{Time-periodic universes}
\author{De-Xing Kong$^{1}$, Kefeng Liu$^{2,3}$ and Ming Shen$^2$}
\affiliation{\\
$^{1}$\!Department of Mathematics,
Zhejiang University, Hangzhou 310027, China\\
$^{2}$\!Center of Mathematical Sciences,
Zhejiang University, Hangzhou 310027, China\\
$^3$\!Department of Mathematics, University of California at Los
Angeles, CA 90095, USA}

\begin{abstract}
In this letter we construct a new time-periodic solution of the
vacuum Einstein's field equations whose Riemann curvature norm takes
the infinity at some points. We show that this solution is
intrinsically time-periodic and describes a time-periodic universe
with the ``black hole". New physical phenomena are investigated and
new singularities are analyzed for this universal model.
\end{abstract}

\pacs{04.20.Jb; 04.20.Dw; 98.80.Jk; 02.30.Jr}

\keywords{Einstein's field equations, time-periodic solution,
time-periodic universe, singularity, black hole.}

\maketitle

{\em 1. Introduction.} The Einstein's field equations are the
fundamental equations in general relativity and play an essential
role in cosmology. The exact solutions of the Einstein's field
equations play a crucial role in the study of general relativity and
cosmology. Typical examples are the Schwarzschild solution and Kerr
solution (see \cite{sch} and \cite{kerr}). Although many interesting
and important solutions have been obtained (see, e.g., \cite{b} and
\cite{skmhh}), there are still several fundamental open problems.
One such problem is {\em if there exists a ``time-periodic"
solution, which contains physical singularities such as black hole,
to the Einstein's field equations}. This letter aims to solves this
problem.

For the evolutionary equations, time-periodic or stationary
solutions correspond to the late time behavior of solutions for a
large class of initial data. In the general theory of relativity,
the time-periodic ``black hole" solutions (if they exist) seem to
provide reasonable candidates for the final state of gravitational
collapse. As pointed out that in \cite{daf}, such solutions can be
defined as those invariant with respect to an isometry of the domain
of outer communications which takes every point to its future, or
more generally, such that points sufficiently close to infinity are
mapped to their future. The study of the periodic solutions to the
Einstein's field equations was initiated in Papapetrou
\cite{pap1}-\cite{pap2}. See also the important paper \cite{gib}.
Dafermos \cite{daf} proved a theorem about the non-existence of
spherically symmetric black-hole space-times with time-periodicity
outside the event horizon, other than Schwarzschild in the vacuum
case and Reissner-Nordstr\"{o}m in the case of electromagnetic
fields and matter sources of a particular kind. This important
result generalizes the ``no-hair" theorem from the static to the
time-periodic case. Up to now, very few results on the
well-posedness for the Einstein's field equations have been
established. In their classical monograph \cite{ck}, Christodoulou
and Klainerman proved the global nonlinear stability of the
Minkowski space for the vacuum Einstein's field equations, i.e.,
they showed the global nonlinear stability of the trivial solution
of the vacuum Einstein's field equations. Lindblad and Rodnianski
\cite{lr} proved the global stability of the Minkowski space for the
vacuum Einstein's field equations in wave coordinate gauge for the
set of restricted data coinciding with the Schwarzschild solution in
the neighborhood of space-like infinity. This work provides a new
and simple approach to the stability problem originally solved by
Christodoulou and Klainerman. In the Ph.D. thesis \cite{z}, Zipser
generalized the result of Christodoulou and Klainerman \cite{ck} to
the Einstein-Maxwell equations. In a series of interesting papers
(see \cite{fksy}-\cite{fsy}), Finster, Kamran, Smoller and Yau
investigated the non-existence of time-periodic solutions of the
Dirac equation, the Einstein-Dirac-Maxwell equations or the
Einstein-Dirac-Yang/Mills equations.

The first exact time-periodic solution of the vacuum Einstein's
field equations was constructed by the authors in \cite{kl}. The
solution presented in \cite{kl} is time-periodic, and describes a
regular space-time, which has vanishing Riemann curvature tensor but
is inhomogenous, anisotropic and not asymptotically flat. In our
recent work \cite{kls}, we construct several kinds of new
time-periodic solutions of the vacuum Einstein's field equations
whose Riemann curvature tensors vanish, keep finite or take the
infinity at some points in these space-times, respectively. However
the norm of Riemann curvature tensors of all these solutions
vanishes. This implies that these solutions essentially describe
regular time-periodic space-times.

In this letter, we construct a new time-periodic solution of the
vacuum Einstein's field equations. For this solution, not only its
Riemann curvature tensor takes the infinity at some points, but also
the norm of the Riemann curvature tensor also go to the infinity at
these points. Therefore, this solution possesses some physical
singularities. We also show that this solution is intrinsically
time-periodic and then can be used to describe a time-periodic
universe with the ``black hole". New physical phenomena are
investigated and new singularities are analyzed for this universal
model. This gives a positive answer of the above open problem.

{\em 2. Time-periodic solution.} Consider the following vacuum
Einstein's field equations
\begin{equation}
G_{\mu\nu}\stackrel{\triangle}{=}R_{\mu\nu}-\frac{1}{2}g_{\mu\nu}R=0,
\end{equation}
or equivalently,
\begin{equation}
R_{\mu\nu}=0,
\end{equation}
where $g_{\mu\nu}\;(\mu,\nu=0,1,2,3)$ is the unknown Lorentzian
metric, $R_{\mu\nu}$ is the Ricci curvature tensor, $R$ is the
scalar curvature and $G_{\mu\nu}$ is the Einstein tensor.

Take $(t,r,\theta,\varphi)$ as the spherical coordinates with
$t\in\mathbb{R},\;r\in [0,\infty),\;\theta\in [0,2\pi),\;\varphi\in
[-\pi/2,\pi/2]$ and let $x^0=t,\;x^1=r,\;x^2=\theta,\;x^3=\varphi$.
In the coordinates $(t,r,\theta,\varphi)$, we consider the metric of
the form
\begin{equation}ds^2=g_{\mu\nu}=u^2dt^2+2qdtdr+2vdtd\varphi -a^2b^2dr^2-a^2d\theta^2,\end{equation}
where $u,\;v,\;a$ are smooth functions of $t,\;r$, and $b,\;q$ are
smooth functions of $t$. It is easy to verify that the determinant
of $(g_{\mu\nu})$ is given by
\begin{equation}g\stackrel{\triangle}{=}\det(g_{\mu\nu})=
-a^4b^2v^2. \end{equation}

For the metric (3), a direct calculation gives
\begin{equation}
R_{02},\;R_{12},\;R_{13},\;R_{23},\;R_{33}=0.
\end{equation}
On the other hand,
\begin{equation}
R_{03}=-\frac{v_{rr}}{2a^2b^2}.
\end{equation}
Noting (2), we have
\begin{equation}
R_{03}=0,\quad {\rm i.e.,}\quad
\frac{v_{rr}}{2a^2b^2}=0.\end{equation} Solving (7) leads to
\begin{equation}
v = cr+d,
\end{equation}
where $c=c(t)$ and $d=d(t)$ are integral functions depending on $t$.
For simplicity, let $d=0$. Then (8) becomes
\begin{equation}
v = cr.
\end{equation}
Substituting (9) into (3) and computing $R_{11},\;R_{22}$ yields
\begin{equation}
R_{11}=-\frac{a^2+2raa_r-2r^2aa_{rr}+2r^2a_r^2}{2r^2a^2}
\end{equation}
and
\begin{equation} R_{22}=\frac{aa_r+raa_{rr}-ra_r^2}{ra^2b^2}.
\end{equation}
Noting (2), we have $R_{22}=0$. Solving it gives
\begin{equation}
a=fr^g,
\end{equation}
where $f,\; g$ are two integral functions depending on $t$. Noting
(2) again yields $R_{11}=0$ and substituting (12) into the equation
$R_{11}=0$ leads to
\begin{equation}
g=-\frac{1}{4}.
\end{equation}
Thus, (12) becomes
\begin{equation}
a=fr^{-\frac{1}{4}}.
\end{equation}
Substituting (9) and (14) into (3) and computing $R_{01}$, we have
\begin{equation}
R_{01}=-\frac{4bf_t+fb_t}{rfb}.
\end{equation}
Noting (2), we have $R_{01}=0$, i.e.,
$$\frac{4bf_t+fb_t}{rfb}=0.$$
Solving this equation gives
\begin{equation}
b=\frac{n}{f^4},
\end{equation}
where $n$ is integral constant. Without loss of generality, we may
assume $n=1$. Then (16) becomes
\begin{equation}
b=\frac{1}{f^4}.
\end{equation}

We now calculate the term $R_{00}$.

Substituting (9), (14) and (17) into (3), by a direct calculation we
obtain
\begin{equation}
R_{00}=-\frac{A}{2r^3cf^2},
\end{equation}
where
\begin{equation}\begin{array}{lll}
A & = & r^{\frac{3}{2}}f^8cu^2-2r^{\frac{5}{2}}f^8cuu_r+
2r^{\frac{7}{2}}f^8cu_r^2+\vspace{2mm}\\
& &
2r^{\frac{7}{2}}f^8cuu_{rr}+4r^3ff_{tt}c-4r^3ff_tc_t-24r^3f_t^2c.\end{array}
\end{equation}
Noting (2) again, we have $R_{00}=0$. Solving this equation gives
\begin{equation}
u^2=4Hr^{\frac{3}{2}}+H_0r\ln r+H_1r,
\end{equation}
where $H_0$ and $H_1$ are two integral functions depending on $t$,
and $H$ is given by
\begin{equation}
H=\frac{24cf_t^2+4c_tff_t-4cff_{tt}}{f^8c}.
\end{equation}

Summarizing the above discussion, we can obtain the following
theorem.

\noindent{\bf Theorem 1} {\em The vacuum Einstein's filed equations
(1) have the following solutions in the coordinates
$(t,r,\theta,\varphi)$
\begin{equation}
ds^2=(dt,dr,d\theta,d\varphi)(g_{\mu\nu})(dt,dr,d\theta,d\varphi)^T,
\end{equation}
where
\begin{equation}(g_{\mu\nu})=\left( \begin{array}{cccc}
4Hr^{\frac{3}{2}}+H_0r\ln r+H_1r & q & 0 & cr\\q & -\frac{1}{f^6\sqrt{r}} & 0 & 0 \\
0& 0 & -\frac{f^2}{\sqrt{r}} & 0\\
cr & 0 & 0 & 0\\
\end{array}\right),\end{equation}
in which $H_0$, $H_1$, $c$, $q$ and $f$ are arbitrary functions of
$t$, and $H$ is defined by (21). $\quad\blacksquare$}

In particular, taking
\begin{equation}
H_0=H_1=0,\quad c=b=\frac{1}{f^4}\end{equation} and
\begin{equation}
f=1+\sin t,\quad q=0,\end{equation} we have

\noindent{\bf Theorem 2} {\em The vacuum Einstein's filed equations
(1) have the following time-periodic solution in the coordinates
$(t,r,\theta,\varphi)$
\begin{equation}
ds^2=(dt,dr,d\theta,d\varphi)(\eta_{\mu\nu})(dt,dr,d\theta,d\varphi)^T,
\end{equation}
where
\begin{equation}\left\{\begin{array}{l}
{\displaystyle \eta_{00}=
\frac{16r^{\frac{3}{2}}(1+\sin t+\cos^2t)}{(1+\sin t)^8}},\vspace{2mm}\\
{\displaystyle \eta_{03}=\frac{r}{(1+\sin t)^4}},\vspace{2mm}\\
{\displaystyle \eta_{11}=-\frac{1}{\sqrt{r}(1+\sin t)^6},}\vspace{2mm}\\
{\displaystyle \eta_{22}=-\frac{(1+\sin t)^2}{\sqrt{r}},}\vspace{2mm}\\
\eta_{01}=\eta_{02}=\eta_{12}=\eta_{13}=\eta_{23}=\eta_{33}=0.
\end{array}\right.\quad\blacksquare\end{equation} }

\noindent{\bf Proof.} By Theorem 1, it is obvious that the metric
(26) is a solution of the vacuum Einstein's filed equations (1). It
suffices to prove the solution (26) is time-periodic. To do so, we
prove that the variable $t$ is a time coordinate.

It is easy to verify that the determinant of $(\eta_{\mu\nu})$ is
given by
\begin{equation}\eta\stackrel{\triangle}{=}\det(\eta_{\mu\nu})=
-\frac{r}{(1+\sin t)^{12}}. \end{equation} Obviously, $t=
2k\pi-\pi/2\;(k\in \mathbb{Z})$ and $r=0$ are the singularities of
the space-time described by (26). A detailed analysis on these
singularities will be given in next section.

When $t\neq 2k\pi-\pi/2\;(k\in \mathbb{Z})$ and $r\neq 0$, it holds
that
$$\eta_{00}=\frac{16r^{\frac{3}{2}}(1+\sin t+\cos^2 t)}{(1+\sin
t)^8}>0,$$
$$
\left|\begin{array}{cccc}
\eta_{00} & \eta_{01} \\
\eta_{01}& \eta_{11}
\end{array}\right|=-\frac{16r(1+\sin t+\cos^2t)}{(1+\sin
t)^{14}}<0,$$
$$\left|\begin{array}{cccc}
\eta_{00} & \eta_{01}&\eta_{02}\\
\eta_{01} & \eta_{11} & \eta_{12} \\
\eta_{20}& \eta_{21} &\eta_{22}
\end{array}\right|=\frac{16\sqrt{r}(1+\sin t+\cos^2 t)}{(1+\sin
t)^{12}}>0$$ and
$$\left|\begin{array}{cccc}
\eta_{00} & \eta_{01} &\eta_{02} & \eta_{03} \\
\eta_{10} & \eta_{11} & \eta_{12} & \eta_{13} \\
\eta_{20} & \eta_{21} & \eta_{22} & \eta_{23} \\
\eta_{30} & \eta_{31} & \eta_{32} & \eta_{33}
\end{array}\right|=-\frac{r}{(1+\sin t)^{12}}<0.$$
This implies that the variable $t$ is a time coordinate. Therefore,
(26) is indeed a time-periodic solution of the vacuum Einstein's
field equations (1). Thus, the proof of Theorem 2 is completed.
$\quad\quad\quad\square$

{\em 3. Singularities.} This section is devoted to the analysis of
singularities of the time-periodic solution (26) of the vacuum
Einstein's field equations.

By direct calculations, the Riemann curvature tensor of (26) reads

\begin{equation}
R_{2121}=\frac{(1+\sin t)^2}{4r^{\frac{5}{2}}},
\end{equation}
\begin{equation}
R_{0101}=\frac{2(1+\sin t)\sin t-2\cos^2 t}{\sqrt{r}(1+\sin t)^{8}},
\end{equation}
\begin{equation}
R_{0221}=\frac{3(1+\sin t)\cos t}{2r^{\frac32}},
\end{equation}
\begin{equation}
R_{0301}=\frac{3\cos t}{2(1+\sin t)^5},
\end{equation}
\begin{equation}
R_{0303}=-\frac{\sqrt{r}}{4(1+\sin t)^2},
\end{equation}
\begin{equation}
R_{0232}=\frac{(1+\sin t)^4}{8r},
\end{equation}
\begin{equation}
R_{0202}=\frac{2(1+\sin t)\sin t+10\cos^2 t}{\sqrt{r}},
\end{equation}
\begin{equation}
R_{0131}=\frac{1}{8r(1+\sin t)^4},
\end{equation}
and the other $R_{\alpha\beta\mu\nu}=0$. Moreover,
\begin{equation}
\mathbf{R}\triangleq
R^{\alpha\beta\gamma\delta}R_{\alpha\beta\gamma\delta}=
\frac{3(1+\sin t)^{12}}{4r^3}.
\end{equation}
Therefore, when $t\neq 2k\pi-\pi/2\;(k\in \mathbb{Z})$ and $r
\rightarrow 0+$, it holds that
\begin{equation}
\mathbf{R}\longrightarrow +\infty.
\end{equation}
Thus, we have

\noindent{\bf Proposition 1} {\em $r=0$ is an essential singular
point. Thus, the solution (26) describes a time-periodic space-time
with a ``black hole". $\quad\blacksquare$}

In particular, when $t= 2k\pi-\pi/2\;(k\in \mathbb{Z})$, we have
$\mathbf{R} = 0$. This implies that the ``black hole" disappears  at
these points.

According to the definition of the event horizon (see \cite{wald}),
the hypersurfaces $t= 2k\pi-\pi/2\;(k\in \mathbb{Z})$ are the event
horizons of the space-time described by (26). Therefore we have

\noindent{\bf Proposition 2} {\em   The solution (26) also contains
non-essential singularities which consist of the hypersurfaces $t=
2k\pi-\pi/2\;\;(k\in \mathbb{Z})$. These hypersurfaces correspond to
the event horizons.}

\noindent{\bf Remark 1} {\em Notice that the curvature tensors are
intrinsic and independent of the choice of the coordinates. It
follows from (29)-(37) that the Lorentzian metric (26) is NOT the
Minkowski metric written in some periodic coordinate system. On the
other hand, it is well known that, by taking a static or stationary
vacuum metric and performing a nontrivial periodic coordinate
transformation one could produce many apparently periodic solutions
- but which are not intrinsically periodic; however the metric (26)
is NOT this case because of (29)-(37). In other words, our solution
(26) is intrinsically time-periodic.}

We now investigate the behavior of the null curves and light-cones
in the space-time (26).

Fixing $\theta$ and $\varphi$, we get the induced metric
$$ds^2=\eta_{00}dt^2+\eta_{11}dr^2.$$
Consider the null curves in the $(t,r)$-plan, which are defined by
$$\eta_{00}dt^2+\eta_{11}dr^2=0.$$
Noting (27) gives
$$\frac{dt}{dr}=\pm\frac{\sqrt{1+\sin t}}{4r\sqrt{2-\sin t}}.$$
Thus, the null curves and light-cones are shown in Figure 1.
\begin{figure}[H]
    \begin{center}
\begin{picture}(258,182)
\thinlines \drawvector{24.0}{12.0}{164.0}{0}{1}
\drawvector{24.0}{12.0}{226.0}{1}{0}
\drawpath{24.0}{150.0}{246.0}{150.0}
\path(28.15,149.12)(29.51,148.8)(30.87,148.5)(32.2,148.16)(33.54,147.83)(34.86,147.5)
\path(34.86,147.5)(36.16,147.14)(37.47,146.8)(38.76,146.42)(40.05,146.05)(41.31,145.67)(42.58,145.28)(43.83,144.88)(45.05,144.49)(46.29,144.08)
\path(46.29,144.08)(47.51,143.66)(48.72,143.22)(49.91,142.8)(51.09,142.35)(52.27,141.89)(53.44,141.44)(54.59,140.97)(55.75,140.5)(56.87,140.0)
\path(56.87,140.0)(58.0,139.52)(59.12,139.02)(60.22,138.5)(61.3,138.0)(62.4,137.47)(63.47,136.94)(64.52,136.39)(65.58,135.85)(66.62,135.3)
\path(66.62,135.3)(67.66,134.72)(68.68,134.16)(69.69,133.58)(70.69,132.99)(71.66,132.39)(72.65,131.78)(73.62,131.17)(74.58,130.55)(75.52,129.92)
\path(75.52,129.92)(76.45,129.3)(77.38,128.64)(78.3,128.0)(79.2,127.33)(80.11,126.66)(80.98,126.0)(81.87,125.3)(82.73,124.61)(83.58,123.91)
\path(83.58,123.91)(84.42,123.21)(85.25,122.5)(86.08,121.77)(86.89,121.03)(87.69,120.3)(88.49,119.55)(89.27,118.8)(90.05,118.02)(90.8,117.25)
\path(90.8,117.25)(91.55,116.47)(92.3,115.69)(93.02,114.9)(93.74,114.08)(94.44,113.27)(95.14,112.45)(95.83,111.62)(96.5,110.8)(97.17,109.94)
\path(97.17,109.94)(97.83,109.09)(98.47,108.23)(99.11,107.37)(99.74,106.5)(100.36,105.61)(100.96,104.72)(101.55,103.81)(102.13,102.91)(102.71,102.0)
\path(102.71,102.0)(103.27,101.06)(103.83,100.13)(104.36,99.19)(104.89,98.25)(105.41,97.3)(105.94,96.33)(106.44,95.36)(106.92,94.37)(107.41,93.38)
\path(107.41,93.38)(107.86,92.4)(108.33,91.38)(108.77,90.37)(109.22,89.36)(109.64,88.33)(110.05,87.3)(110.47,86.25)(110.86,85.19)(111.25,84.13)
\path(111.25,84.13)(111.63,83.06)(111.99,82.0)(112.0,82.0)
\path(112.0,82.0)(112.0,82.0)(112.52,83.23)(113.07,84.44)(113.63,85.66)(114.19,86.86)(114.78,88.05)(115.39,89.23)(116.02,90.38)(116.64,91.54)
\path(116.64,91.54)(117.3,92.69)(117.97,93.8)(118.66,94.93)(119.36,96.04)(120.07,97.12)(120.8,98.22)(121.55,99.29)(122.3,100.34)(123.08,101.4)
\path(123.08,101.4)(123.88,102.44)(124.69,103.45)(125.52,104.48)(126.35,105.48)(127.21,106.47)(128.08,107.44)(128.97,108.41)(129.86,109.37)(130.78,110.3)
\path(130.78,110.3)(131.72,111.25)(132.66,112.16)(133.63,113.08)(134.61,113.97)(135.61,114.86)(136.61,115.73)(137.64,116.6)(138.69,117.44)(139.75,118.28)
\path(139.75,118.28)(140.82,119.11)(141.91,119.92)(143.02,120.74)(144.13,121.52)(145.27,122.3)(146.42,123.08)(147.58,123.83)(148.77,124.58)(149.97,125.33)
\path(149.97,125.33)(151.19,126.05)(152.41,126.75)(153.66,127.46)(154.92,128.14)(156.19,128.83)(157.49,129.49)(158.8,130.14)(160.13,130.78)(161.47,131.41)
\path(161.47,131.41)(162.82,132.03)(164.19,132.64)(165.58,133.25)(166.97,133.83)(168.38,134.39)(169.83,134.97)(171.27,135.5)(172.74,136.05)(174.22,136.58)
\path(174.22,136.58)(175.71,137.08)(177.22,137.6)(178.75,138.08)(180.28,138.57)(181.85,139.03)(183.41,139.5)(185.0,139.94)(186.61,140.36)(188.22,140.8)
\path(188.22,140.8)(189.86,141.21)(191.52,141.61)(193.19,141.99)(194.86,142.36)(196.57,142.72)(198.27,143.08)(200.0,143.42)(201.75,143.75)(203.5,144.07)
\path(203.5,144.07)(205.28,144.38)(207.08,144.67)(208.88,144.96)(210.71,145.22)(212.54,145.49)(214.39,145.74)(216.26,145.97)(218.16,146.19)(220.05,146.41)
\path(220.05,146.41)(221.97,146.61)(223.91,146.8)(225.85,146.97)(227.82,147.14)(229.79,147.3)(231.79,147.44)(233.79,147.58)(235.82,147.69)(237.86,147.8)
\path(237.86,147.8)(239.92,147.91)(241.98,148.0)(242.0,148.0)
\path(112.0,82.0)(112.0,82.0)(111.5,80.91)(111.02,79.84)(110.52,78.79)(110.0,77.73)(109.5,76.69)(108.97,75.65)(108.44,74.62)(107.89,73.59)
\path(107.89,73.59)(107.35,72.58)(106.8,71.58)(106.22,70.56)(105.66,69.58)(105.08,68.59)(104.49,67.62)(103.88,66.65)(103.28,65.69)(102.67,64.73)
\path(102.67,64.73)(102.05,63.79)(101.42,62.84)(100.8,61.91)(100.14,60.98)(99.5,60.06)(98.83,59.16)(98.16,58.26)(97.5,57.37)(96.8,56.48)
\path(96.8,56.48)(96.11,55.61)(95.41,54.73)(94.71,53.87)(94.0,53.01)(93.27,52.16)(92.53,51.33)(91.8,50.49)(91.05,49.66)(90.3,48.84)
\path(90.3,48.84)(89.52,48.04)(88.75,47.24)(87.97,46.43)(87.19,45.65)(86.39,44.88)(85.58,44.09)(84.77,43.34)(83.95,42.58)(83.12,41.83)
\path(83.12,41.83)(82.3,41.09)(81.44,40.36)(80.59,39.63)(79.73,38.9)(78.87,38.2)(78.0,37.5)(77.11,36.79)(76.22,36.11)(75.31,35.43)
\path(75.31,35.43)(74.41,34.75)(73.5,34.09)(72.56,33.43)(71.63,32.77)(70.69,32.13)(69.75,31.5)(68.8,30.88)(67.83,30.25)(66.86,29.63)
\path(66.86,29.63)(65.87,29.04)(64.88,28.43)(63.9,27.84)(62.88,27.27)(61.87,26.68)(60.86,26.13)(59.83,25.56)(58.8,25.02)(57.75,24.47)
\path(57.75,24.47)(56.69,23.93)(55.63,23.4)(54.56,22.88)(53.5,22.36)(52.41,21.86)(51.31,21.36)(50.22,20.86)(49.11,20.38)(48.0,19.92)
\path(48.0,19.92)(46.87,19.45)(45.73,18.98)(44.59,18.53)(43.44,18.09)(42.3,17.64)(41.12,17.21)(39.95,16.79)(38.77,16.37)(37.58,15.96)
\path(37.58,15.96)(36.4,15.57)(35.19,15.18)(33.98,14.79)(32.76,14.42)(31.52,14.04)(30.3,13.69)(29.05,13.34)(27.8,12.98)(26.54,12.64)
\path(26.54,12.64)
\path(112.0,82.0)(112.0,82.0)(112.36,80.8)(112.75,79.62)(113.16,78.44)(113.58,77.27)(114.03,76.12)(114.5,74.98)(115.0,73.84)(115.5,72.73)
\path(115.5,72.73)(116.02,71.62)(116.58,70.51)(117.13,69.41)(117.72,68.33)(118.33,67.26)(118.96,66.2)(119.6,65.16)(120.25,64.12)(120.94,63.09)
\path(120.94,63.09)(121.64,62.08)(122.36,61.06)(123.11,60.08)(123.88,59.08)(124.66,58.11)(125.46,57.15)(126.27,56.19)(127.11,55.25)(127.97,54.3)
\path(127.97,54.3)(128.86,53.38)(129.75,52.47)(130.67,51.56)(131.61,50.68)(132.57,49.79)(133.55,48.91)(134.55,48.06)(135.55,47.2)(136.6,46.36)
\path(136.6,46.36)(137.66,45.52)(138.72,44.7)(139.83,43.9)(140.94,43.09)(142.07,42.31)(143.22,41.54)(144.39,40.77)(145.6,40.0)(146.8,39.25)
\path(146.8,39.25)(148.03,38.52)(149.28,37.79)(150.55,37.08)(151.85,36.38)(153.16,35.68)(154.49,35.0)(155.83,34.31)(157.21,33.65)(158.6,33.0)
\path(158.6,33.0)(160.0,32.36)(161.44,31.72)(162.88,31.09)(164.36,30.49)(165.83,29.88)(167.35,29.29)(168.86,28.72)(170.41,28.13)(171.99,27.58)
\path(171.99,27.58)(173.57,27.02)(175.17,26.49)(176.8,25.97)(178.44,25.45)(180.11,24.93)(181.78,24.43)(183.49,23.95)(185.21,23.47)(186.96,23.0)
\path(186.96,23.0)(188.72,22.54)(190.5,22.11)(192.3,21.66)(194.11,21.25)(195.96,20.83)(197.82,20.43)(199.69,20.02)(201.6,19.64)(203.5,19.28)
\path(203.5,19.28)(205.44,18.9)(207.41,18.55)(209.38,18.21)(211.38,17.88)(213.39,17.56)(215.44,17.25)(217.48,16.95)(219.57,16.65)(221.66,16.37)
\path(221.66,16.37)(223.76,16.12)(225.91,15.86)(228.05,15.61)(230.23,15.37)(232.42,15.13)(234.64,14.93)(236.86,14.71)(239.11,14.52)(241.39,14.34)
\path(241.39,14.34)(243.67,14.15)
\drawellipse{106.0}{108.0}{0.0}{0.0}{}
\drawellipse{102.0}{104.0}{0.0}{0.0}{}
\drawellipse{102.0}{104.0}{0.0}{0.0}{}
\drawellipse{113.0}{102.0}{22.0}{4.0}{}
\drawellipse{111.0}{61.0}{22.0}{6.0}{}
\drawcenteredtext{13.0}{150.0}{$\frac{3\pi}{2}$}
\drawcenteredtext{16.0}{174.0}{$t$}
\drawcenteredtext{252.0}{4.0}{$r$}
\drawcenteredtext{13.0}{12.0}{$-\frac{\pi}{2}$}
\end{picture}
\caption{Null curves and light-cones in the domains
$-\pi/2<t<3\pi/2$.}
    \end{center}
\end{figure}
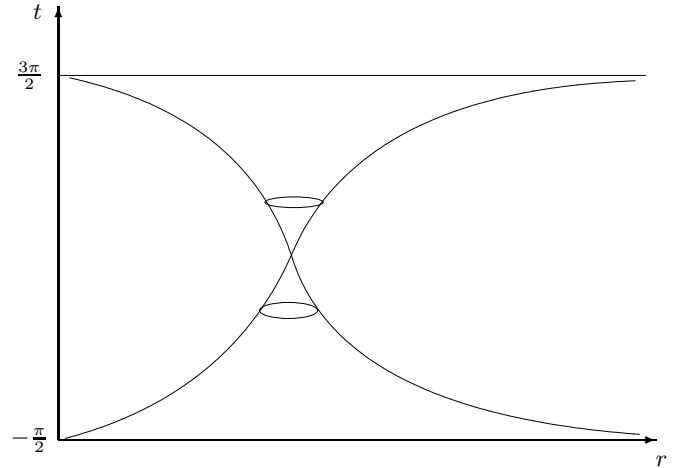

We next study the geometric behavior of the $t$-slices.

For any fixed $t\in \mathbb{R}$, it follows from (26) that the
induced metric of the $t$-slice reads
\begin{equation}
ds^2= -\frac{1}{\sqrt{r}(1+\sin t)^2}[dr^2+(1+\sin t)^8d\theta^2].
\end{equation}
As mentioned before, the hypersurfaces $t = 2k\pi-\pi/2\;\;(k\in
\mathbb{Z})$ are singularities of the space-time described by (26),
while, when $t\neq 2k\pi-\pi/2\;\;(k\in \mathbb{Z})$, the $t$-slice
is a three-dimensional cone-like manifold centered at $r=\infty$.

{\em 4. Summary and discussion.} In our previous works
\cite{kl}-\cite{kls}, we have constructed three kinds of new
time-periodic solutions of the vacuum Einstein's field equations:
the regular time-periodic solution with vanishing Riemann curvature
tensor, the regular time-periodic solution with finite Riemann
curvature tensor and the time-periodic solution with physical
singularities. However, the norm of the Riemann curvature tensors of
all these solutions vanishes, therefore these solutions essentially
describe some regular time-periodic space-times, these space-times
contain some non-physical singularities, but no physical
singularity.

In this letter we construct a new time-periodic solution (26) of the
vacuum Einstein's field equations. The norm of the Riemann curvature
tensor of the metric (26) goes to the infinity when $r$ tends to
zero. Therefore $r=0$ is a physical singularity of the space-time
described by (26), which is named as ``black hole" in this letter.
This solution also contains some non-essential singularities which
consist of the hypersurfaces $t= 2k\pi-\pi/2\;\;(k\in \mathbb{Z})$.
In particular, by (29)-(37) we observe that the metric (26) is
impossible to be the Minkowski metric written in some periodic
coordinate system; on the other hand, by taking a static or
stationary vacuum metric and performing a nontrivial periodic
coordinate transformation one could NOT produce the metric (26),
this is to say, the solution (26) is intrinsically time-periodic. As
a corollary, we would like to point out that the space-time (26) has
a time-periodic time-like Killing vector field. Moreover, new
physical phenomena have been investigated for the time-periodic
universal model characterized by (26). Consequently, the solution
(26) solves the long-time open problem mentioned at the first
paragraph in Section 1, and more applications of this new space-time
in modern cosmology and general relativity can be expected.

The work of Kong was supported in part by the NSF of China (Grant
No. 10671124) and the Qiu-Shi Professor Fellowship from Zhejiang
University, China; the work of Liu was supported by the NSF and NSF
of China.

\end{document}